\documentclass[preprint,trackchanges]{aastex631}
\usepackage [latin1]{inputenc}
\hypersetup{linkcolor=blue,citecolor=blue,filecolor=cyan,urlcolor=blue}
\shorttitle{Funnel Prominence Formation}
\shortauthors{Yang et al.}

\begin{document}

\title{Observations of the Formation and Disappearance of a Funnel Prominence}

\correspondingauthor{Bo Yang}
\email{boyang@ynao.ac.cn}

\author{Bo Yang}
\affil{ Yunnan Observatories, Chinese Academy of Sciences, 396 Yangfangwang, Guandu District, Kunming, 650216, People's Republic of China}
\affil{ Yunnan Key Laboratory of Solar Physics and Space Science, Kunming 650011, People¡¯s Republic of China}

\author{Jiayan Yang}
\affil{ Yunnan Observatories, Chinese Academy of Sciences, 396 Yangfangwang, Guandu District, Kunming, 650216, People's Republic of China}
\affil{ Yunnan Key Laboratory of Solar Physics and Space Science, Kunming 650011, People¡¯s Republic of China}

\author{Yi Bi}
\affil{ Yunnan Observatories, Chinese Academy of Sciences, 396 Yangfangwang, Guandu District, Kunming, 650216, People's Republic of China}
\affil{ Yunnan Key Laboratory of Solar Physics and Space Science, Kunming 650011, People¡¯s Republic of China}

\author{Junchao Hong}
\affil{ Yunnan Observatories, Chinese Academy of Sciences, 396 Yangfangwang, Guandu District, Kunming, 650216, People's Republic of China}
\affil{ Yunnan Key Laboratory of Solar Physics and Space Science, Kunming 650011, People¡¯s Republic of China}



\begin{abstract}
We present an observational study of the formation and disappearance of a funnel prominence.
Before the funnel prominence formed, cool materials from the top of a preexisting polar crown prominence
flowed along saddle-shaped coronal loops to their base, forming a smaller prominence.
Meanwhile, the saddle-shaped coronal loops gradually rose, and U-shaped coronal loops, termed prominence horns, began to appear along with a coronal cavity.
Afterwards, a cool column emerged from the chromosphere, rose vertically into the corona,
and then moved laterally to be  transported  into the U-shaped coronal loops.
The formed prominence slid into the chromosphere, while the U-shaped coronal loops and the coronal cavity became more pronounced.
As cool materials accumulated at the base of the U-shaped coronal loops, these loops underwent a significant descent 
and a V-shaped structure appeared at the base of the cool materials, indicating that the U-shaped coronal loops may be dragged down to sag.
Subsequently, cool materials from the V-shaped structure continued to flow almost vertically toward the  chromosphere, forming the funnel prominence.
The vertical downflows might be produced by magnetic reconnection within or between the sagging field lines.
Due to persistent vertical downflows, the U-shaped coronal loops  were lifted up and prominence materials followed along inclined coronal loops towards the chromosphere,
causing the funnel prominence to disappear.  
Our observations suggest that chromospheric plasma transported into a coronal cavity and then drained out via vertical downflows
can form a funnel prominence.
\end{abstract}


\section{Introduction}  \label{sec:intro}

Solar prominences are relatively cool, dense, plasma structures suspended in the much hotter, tenuous corona.
They are commonly found in the solar atmosphere and are distributed from the active latitudes all the way to the polar crown,
implying that prominences are a manifestation of a common physical process occurring in the solar atmosphere.
To understand how prominences are supported, where and why they form, as well as how the overdense prominence materials and magnetic fields dynamically interact
have long been the central problems in prominence physics \citep[see reviews by][]{mar98,mac10, ber14,pare14,gibson18,chen20b}.
The study of prominence formation  provides an opportunity to investigate these issues and may aid in a better understanding  of the evolution of magnetic fields
and the mass cycles in the solar atmosphere.

Prominences are often surrounded by a low-density coronal structure known as a coronal cavity \citep{gibson10,gibson18}. This is especially true with polar crown prominences.
Coronal cavities are observed as circular, half-circular, or elliptical darkened regions above the solar limb in white light \citep{gibson06b} and extreme ultraviolet (EUV) coronal images \citep{karna15a,karna15b}, 
particularly visible in the 193 \AA\ passband of the Atmospheric Imaging Assembly \citep[AIA;][]{lem12} on board the \emph{Solar Dynamics Observatory} \citep[\emph{SDO};][]{pes12}. 
They are the limb counterpart to the filament channel on the disk \citep{gibson06b,karna17}
and are in general believed to be representative of magnetic flux ropes made up of twisted magnetic field lines \citep{gibson06a,bak13,gibson18}.
When viewed down the axis of the magnetic flux ropes, the helical field lines on the interior of the magnetic flux ropes form an elliptical structure.
The magnetic fields in the magnetic flux ropes on the Sun are under-dense due to the force balance with surrounding coronal regions and thus appear darker
than the surrounding corona plasma \citep{gibson06a,gibson15}.  In the \emph{SDO}/AIA 171 \AA\ observation, coronal plasma is usually shaped by magnetic field lines with a negative curvature, 
thus appearing as U-shaped structures sitting at the bottom of a coronal cavity \citep{reg11,wang16}.
The U-shaped structures, dubbed prominence horns \citep{ber12,sch13,wang16}, emanate from the prominence and extend into the coronal cavity.

It is widely accepted that prominence magnetic fields support the cold dense prominence materials against gravity and insulate them from the surrounding hot corona.
The coronal cavity reflects the fact that prominences have a flux rope topology with dips in the magnetic field lines \citep{gibson06b,gibson15}.
Direct measurement of prominence magnetic fields \citep{ler89,oro14,schm14}, theoretical models  \citep{ks57,kr74,van89,aul98,dev00},
and observations \citep{cheng14,yan14,yang14,fil15,wang15,ouyang17}
have indicated that magnetic dips of twisted magnetic flux ropes or sheared arcades are believed to be suitable for supporting channel prominences,
which have spines and are usually formed above the polarity inversion lines (PILs),
a line separating two large regions of opposite vertical components of the photospheric magnetic fields \citep{mar98}.
The \emph{SDO}/AIA observations revealed a potentially new type of prominence, termed ``funnel prominences''  \citep{ber14,liu14}.
They feature a characteristic inverted cone structure \citep{kle72} and share some morphological similarities to quiescent prominences, 
such as the conical shapes at the tops of funnels and the prominence horns \citep{ber12,liu14}.
However, whether funnel prominences are surrounded by a coronal cavity and must be associated with PILs remain heretofore unknown \citep{ber14,liu14}.
In earlier ground-based observations, funnel prominences and the so called ``coronal spiders'' \citep{all98} were categorized  as coronal cloud prominences  \citep{martin15}.
Coronal spiders appear closer in character to coronal rain events, in which plasma condenses high in the corona 
and then falls under gravitational attraction along magnetic field lines to the chromosphere \citep{anto12}. 
In contrast, downflows in funnel prominences are constrained to narrow conical regions, and they flow along meandering paths down to the chromosphere at much lower speeds than free-fall speeds, 
suggestive of cross-field mass transport \citep{liu12,liu14}. A basic question of funnel prominences is what type of magnetic field configuration supports them.
Some previous observational studies have provided clues that magnetic dips of coronal arcades \citep{kle72,iva77,liu12,kep14,sch16,li18b,fil23},
separatrix surfaces between adjoining coronal loop systems \citep{lin06,martin15},
and the funnel-like open coronal magnetic fields \citep{pan19} may be the sites for the formation and accumulation of funnel prominence mass.

At present, it is generally believed that the material of channel prominences originates from  the solar photosphere or chromosphere \citep{spicer98,song17}.
Much progress has been made in revealing the mechanisms that account for the transport of the photospheric or  chromospheric material into the corona to form channel prominences.
Chromospheric material can be directly injected into a filament channel through intermittent magnetic reconnection in the form of surges or jets \citep{wang99,li16,wang18,shen19,yang19a,yang19b,li23}.
Besides, photospheric or  chromospheric materials could be lifted into the corona by the emergence of a helical flux rope \citep{oka08,oka10,yan17}.
Numerical simulations have illustrated that chromospheric material can be heated and evaporated into the corona via localized heating
concentrated exclusively at the chromospheric footpoints of the prominence, and then condensed as cool prominence material \citep{ant91,kar08,luna12,kan15,xia16,zhou20,huang21}.
This is the so-called evaporation-condensation model and has recently been confirmed by our observations \citep{yang21}, as well as those of \citet{li21}.
More recently, observations have demonstrated that the formation of prominences may not be a single mechanism,
but may be under the combined action of several mechanisms \citep{huang23,li23}.

The mass source and transport mechanisms of  funnel prominences still remain questionable.
A handful of observations have presented that in situ condensation of coronal hot plasmas in magnetic dips can result in the formation of funnel prominences \citep{liu12,sch16,fil23}.
By implementing a 2.5D numerical simulation, \citet{kep14} found that continuous condensation of  hot plasma in magnetic dips of arcaded field lines
through ongoing evaporation-condensation process can lead to the formation of funnel prominences.  In addition to the chromospheric evaporation,
the injected hot jets \citep{li23}, emerging magnetic bubbles and plumes \citep{ber11} might all be potential sources of the hot plasma.
Furthermore, several authors have proposed that the mass source of funnel prominences could even come from the mass falling back to the Sun
in the wake of the eruption in the surrounding atmosphere \citep{lin06,pan19}.

High spatiotemporal images of the prominences have revealed a wealth of fine, highly dynamic phenomena within the prominences, such as rising bubbles and plumes \citep{ber11,shen15},
vortexes \citep{li18a}, descending knots \citep{chae08,chae10,bi20}, and dynamic vertical threads \citep{chae08,chae10}, etc.
It is not fully force-free inside the prominence \citep{anz07}. Gravity forces play an essential role in the dynamic evolution of the prominence \citep{gibson18}.
The descending knots and the dynamic vertical threads, which follow along the direction of gravity
and may imply the prominence plasma to slip across magnetic field lines towards the solar surface, are gravity-driven flows.
\citet{chae08} found that a pattern of persistent horizontal flows originated from a chromospheric region and rose to corona to form descending knots and vertical threads.
They suggested that the horizontal flows provide mass to sag the initially horizontal magnetic field lines,
resulting in a vertical stack of magnetic dips suitable for the support and development of vertical threads or bright descending knots.
\citet{liu12} observed that continuous condensed prominence mass gradually formed a V-shaped structure,
and sustained mass drainage originated from the V-shaped structure and orientated along vertical direction as dynamic vertical threads.
A number of theories and simulations have managed to investigate how the addition and accumulation of prominence material alters the structure of the coronal magnetic field
and then produces the descending knots and the dynamic vertical threads.
They found that as the prominence material continuous to be added and accumulated in the magnetic dips during the prominence formation,
the accumulated prominence material can weigh down the dipped field lines, causing them to sag with its weight and compress them at the base of the prominence \citep{hill13,terr15}.
As a result, magnetic reconnection may take place within or between the sagging, dipped field lines to form
the descending knots and the dynamic vertical threads \citep{ler80,pet05,chae10,hill12,low12b,kep14,jenk21}.
It is noteworthy that the initial compression of the magnetic field by the heavy prominence should create a large but finite current between the flux surfaces,
which would result in the  spontaneous breakdown in the frozen-in condition, allowing material to pass from one field line to another in the direction of gravity \citep{low12a}.
Nevertheless,  the cross-field diffusion of neutral matter should also play an important role in the mass loss and dynamics of the prominence \citep{gil02}.
Although previous studies have highlighted the important role of gravity in the dynamic evolution of prominences,  
the detailed process by which heavy prominence material interacts with magnetic fields has not yet been directly observed.

In this paper, we present observations of a funnel prominence formed by transporting chromospheric materials into a coronal cavity.
This provides us with a good opportunity to investigate in detail the dynamical interaction process between the heavy prominence materials and the magnetic fields supporting them.
In particular, this observation has the potential to enhance our understanding of the formation, support, and disappearance of funnel prominences and dynamic vertical threads.

\section{Observations}
The event was well-observed by the \emph{SDO}/AIA,  which takes the full-disk images of the Sun in seven extreme ultraviolet (EUV)
and two ultraviolet (UV) wavelengths with a pixel size of 0$\arcsec$.6 at a high cadence of up to 12s.
Here, the Level 1.5 images centered at 304 \AA \ (\ion{He}{2}, 0.05 MK) were adopted to study  the transport of chromospheric materials into the coronal cavity
and the subsequent formation of the funnel prominence.
The Level 1.5 images centered at 171 \AA\ (Fe {\sc ix}, 0.6 MK ) and 193 \AA\ (Fe {\sc xii}, 1.3 MK and Fe {\sc xxiv}, 20 MK)
were used to investigate the magnetic field configuration of the funnel prominence and the evolution of the horn-like structures associated with the funnel prominence.
In addition, this event was also monitored with routine observations from the Global Oscillation Network Group \citep[GONG;][]{har96} in $H_{\alpha}$.
The GONG $H_{\alpha}$ images have a pixel size of 1$\arcsec$.0 and a cadence of 1 minute.
The AIA data used in this study were taken between 19:59 UT on 2015 January 2 and 20:00 UT on 2015 January 3,
and the GONG data were taken between 07:30 UT and 20:00 UT on 2015 January 3.
They were processed by using standard software programs in the Solar Software (SSW),
and aligned by differentially rotating to the reference time of 18:00 UT on 2015 January 3.

\section{Results}
\subsection{Disappearance of a Preexisting Prominence and Formation of a New Smaller One}
The funnel prominence in question was formed on 2015 January 3. Before the funnel prominence formed,
there was a small polar crown prominence seen beyond the northwest limb of the Sun.
The funnel prominence formed to the north of the preexisting prominence.
Figure 1({\it a}) illustrates the general appearance of the preexisting prominence at 20:00 UT on 2015 January 2.
From about 20:00 UT on 2015 January 2 to 05:00 UT on 2015 January 3, the preexisting prominence is disturbed.
It underwent a dynamic evolution process and its main body disappeared.
Persistent prominence materials drained from its top down to the solar surface along curved paths (Figure 1({\it b} \sbond {\it d})),
reminiscent of the coronal rain caused by the falling of the prominence material \citep{sch16,anto22,chen22}.
By about 00:00 UT on January 3, the falling prominence materials split into two branches, the lower one sliding along the original path,
and the upper one moving along a curved path that did not fall back to the solar surface (Figure 1({\it d})).
In the meantime, saddle-shaped coronal loops, which are highly consistent with the path of the upper branch, appeared (Figure 1({\it e}) and Figure 1({\it h})).
The draining prominence materials traced out of these saddle-shaped coronal loops,
along which they traveled and accumulated to their base to form a new smaller prominence  (Figure 1({\it e} \sbond {\it g})).
With the draining of the prominence materials and the formation of the new prominence,
the left half of the saddle-shaped coronal loops, which correspond to the top part of the preexisting prominence,
gradually rose and moved closer to the newly formed prominence (see the associated animation).
Subsequently, U-shaped coronal loops began to appear, which emanate from the newly formed prominence and extend into the corona (Figure 1({\it g})).
The preexisting prominence finally disappeared, leaving only a small fraction of the prominence materials suspended in the corona.
We speculate that the accumulated prominence materials may weigh down and drag the saddle-shaped coronal loops, causing them to sag.
Moreover, as the prominence materials are drained,  the force balance on the saddle-shaped loops changes and this may cause them to appear to rise.
The U-shaped coronal loops are reminiscent of the prominence horns \citep{ber12,sch13,wang16},
which sit on top of a prominence and usually depict the boundary of a coronal cavity.
The rise of the saddle-shaped loops could also be due to rotation of a coronal cavity flux rope over the limb, 
and  the appearance of the U-shaped coronal loops might be the result of the rotation of the coronal cavity into a favorable viewing angle.
These observations indicate that the prominence materials are supported in the magnetic dips of  the sagged coronal magnetic field lines,
and they can be transferred from one part of the prominence to another to form a new prominence.

\subsection{Chromospheric Materials Transported into a Coronal Cavity and Formation of a Funnel Prominence}
Figure 2 shows  the transport of chromospheric materials into a coronal cavity to form the funnel prominence.
Immediately after the formation of the small prominence, a cool column emerged from the solar chromosphere below the suspended residual prominence materials
and rose into the corona ( see Figure 1({\it g}) and Figure 2({\it a} \sbond {\it b})).
The cool column rotated clockwise as it rose (see the associated animation), which shows some similarities to the untwisting motion of the erupted minifilaments.
With the emergence and rise of the cool column, the residual prominence materials gradually rose and moved laterally towards the newly formed small prominence.
The rising of the cool column and the residual prominence materials are clearly shown on a spacetime plot (Figure 3({\it c})),
which along a dotted line,``AB,'' in Figure 2({\it b}), was constructed from AIA 304 \AA\ images.
The top of the cool column rose with a mean velocity of about 13.2 km s$^{-1}$, which is higher than the rising motion of the emerging flux observed by \citet{oka10}.
The rising cool column might be a rising minifilament or a cool magnetic flux rope with a negative, left-hand twist.

The rise of the cool column lasted approximately 1.5 hr. By about 06:40 UT on  2015 January 3, the cool column reached a projected height of about 32 Mm.
It then moved laterally towards the previously formed prominence. At the same time, the U-shaped coronal loops, more precisely termed prominence horns, 
gradually became more pronounced (Figure 2({\it c} \sbond {\it e})). 
Of particular note is that a coronal cavity also gradually appeared above the horn-like structure (see Figure 2({\it c} \sbond {\it i}) and Figure 3({\it a} \sbond {\it b})).
These observations may further support our conjecture that the appearance of the horn-like structure and the coronal cavity might be caused by 
the rotation of a coronal cavity flux rope structure into a favorable viewing angle.
Eventually, most of the material in the cool column was transferred into the coronal cavity (Figure 2({\it d})).
The process of the cool column moving laterally and being transported into the coronal cavity can be clearly shown on a spacetime plot (Figure 3({\it d})),
which along a dotted line,``CD,'' in Figure 2({\it c}), was constructed from AIA 304 \AA\ images.
The cool column moved laterally and  was transferred into the coronal cavity at a speed of about 20.4 km s$^{-1}$.
It is possible to be disturbed by the mass transport, the previously formed prominence slid to the solar surface and disappeared (Figure 2({\it b} \sbond {\it d})).
Subsequently, it is clear that the cool materials at the bottom of the coronal cavity underwent a to-and-fro movement
during 08:00 UT-09:30 UT on 2015 January 3 (see Figure 2({\it d} \sbond {\it f}) and associated animation).
Thereafter,  cool materials accumulated at the bottom of the coronal cavity and then started to extend almost vertically toward the solar surface,
forming a funnel prominence with a typical inverted cone structure pointing sunward and a distinct horn-like structure attached to its top 
and connected it to a coronal cavity (Figure 2({\it g} \sbond {\it i})).

It is evident from the AIA 171 \AA\ difference images (Figure 3({\it a} \sbond {\it b})) that the U-shaped coronal loops
gradually sunk down during the to-and-fro movement of the cool materials.
This suggests that the cool materials transported into the coronal cavity may significantly deform the magnetic fields of the coronal cavity.
It is likely that the cool materials in the coronal cavity may drag the magnetic fields of the coronal cavity down and compress those magnetic fields 
that lie directly below the cool materials, causing the associated magnetic fields to sink.

The most notable characteristic observed during the formation of the funnel prominence was the development of a V-shaped structure
at the base of the accumulated cool materials after the to-and-fro movement of the cool materials (Figure 4({\it a1} \sbond {\it a2})).
The V-shaped structure, which hints at the presence of magnetic dips, may be created by the sagging of the magnetic fields of the coronal cavity by the accumulated cool materials.
Subsequently, persistent cool materials stemmed from this structure and slid towards the solar chromosphere (Figure 4({\it a2} \sbond {\it a4}) and Figure 4({\it b2} \sbond {\it b4})).
To track the dynamic evolution of the sliding cool materials, a spacetime plot was constructed along slice ``EF'' in Figure 4({\it a5}) using AIA 304 \AA\ images,
and the result was provided in Figure 4({\it e}).  We find that the projected velocity of the sliding cool materials reached up to about 40.7 km s$^{-1}$,
which is in line with previous observations \citep{chae08,chae10,liu12} and simulations \citep{kep14,jenk21}.
Accompanied by the persistent cool material sliding towards the solar chromosphere,
the inverted cone structure of the funnel prominence gradually formed  (Figure 4({\it a4} \sbond {\it a5})).
Furthermore, the GONG $H_{\alpha}$ observations show that the intensity at the central part of the prominence
also gradually increased towards the solar chromosphere (Figure 4({\it b1} \sbond {\it b5})).
The formed funnel prominence showed up as a vertical absorption thread on the AIA 193  \AA\ image  (Figure 4({\it c})).
The formed V-shaped structure, the sustained sliding of cool materials from its underside,
and the formation of the vertical absorption thread might be consistent with the scenario of cross-field mass transport
caused by magnetic reconnection within or between the sagging, dipped field lines \citep{ler80,pet05,chae10,liu12,low12b,kep14,jenk21}.
It is worth mentioning that the U-shaped coronal loops rose up significantly
in the course of the cool material sliding toward the chromosphere (Figure 4({\it d})).
Magnetic reconnection within or between the sagging, dipped field lines may alter the local magnetic field topology of the prominence
and lead to significant mass loss due to the vertical downflows.
This can result in the upward magnetic tension force that exceeds the gravity force, causing the U-shaped coronal loops to rise.

\subsection{Disappearance of the Funnel Prominence by Persistent Mass Drainage}
The funnel prominence lived for about seven hours and eventually disappeared. Figure 5 shows the detailed process of the disappearance of the funnel prominence.
As the cool materials continued to slide from the bottom of the prominence towards the chromosphere, from about 12:20 UT on  2015 January 3,
a large amount of cool materials flowed from the prominence along well-defined curved paths towards the solar surface (Figure 5({\it a} \sbond {\it c})).
By about 18:40 UT on  2015 January 3, the funnel prominence  completely disappeared (Figure 5({\it d})).
Scrutinizing the AIA 171  \AA\ negative images, it is found that the U-shaped coronal loops continued to rise, gradually became flatter,
and finally  evolved into inclined coronal loops (Figure 5({\it e} \sbond {\it h})).
Comparing the AIA 304 \AA\ and  171 \AA\ observations, it is obvious that the prominence materials
are drained towards the solar chromosphere along these inclined coronal loops.
On the basis of our observations, we conjecture that the disappearance of the funnel prominence
may be caused by the drainage of the prominence materials along well-defined curved coronal loops,
which may be initiated by the persistent sliding of cool materials from the base of the prominence towards the chromosphere.
Owing to the persistent sliding of cool materials from the base of the prominence towards the chromosphere, the balance between magnetic tension and gravitational forces is broken down.
As a result, the surplus magnetic tension force may cause the U-shaped coronal loops and the cool materials trapped in them to be lifted,
then the U-shaped coronal loops become flatter and eventually involve into inclined coronal loops.
At the same time,  the prominence materials drain along the inclined coronal loops towards the solar surface, leading it to disappear completely.

\section{Conclusion and Discussion}
Employing \emph{SDO}/AIA EUV and GONG $H_{\alpha}$ observations, the formation  and disappearance of a funnel prominence is investigated in detail.
Prior to the formation of the funnel prominence, we find that cool materials from the top of a preexisting polar crown prominence
flowed partly along a curved path down to the solar chromosphere and partly along saddle-shaped coronal loops to their base.
This results in the disappearance of the preexisting prominence and the simultaneous formation of a new, smaller one.
The saddle-shaped coronal loops slowly raised and U-shaped coronal loops associated with a coronal cavity gradually began to appear.
Immediately after the formation of the small prominence, a cool column emerged from the solar chromosphere and was lifted into the corona.
Upon reaching a projected height of approximately 32 Mm above the solar surface, the cool column moved laterally.
Cool materials carried by the cool column then were transported into the U-shaped coronal loops and accumulated at their base. 
The formed prominence slid into the solar chromosphere and disappeared.
In the meantime, the U-shaped coronal loops and the coronal cavity  became more pronounced.
The U-shaped coronal loops are in line with prominence horns, which depict the boundary of the coronal cavity.
The rise of the saddle-shaped loops could be due to rotation of a coronal cavity flux rope over the limb, 
and the appearance of the U-shaped coronal loops associated with the coronal cavity might be the result of the rotation of the coronal cavity flux rope into a favorable viewing angle.
Subsequently, a V-shaped structure formed at the base of the accumulated cool materials, from which persistent cool materials slid almost vertically towards the solar chromosphere,
forming a funnel prominence with a typical inverted cone structure and a distinct horn-like structure emanating from the prominence and extending into a coronal cavity.
Due to the vertical drainage of the prominence materials, a large amount of prominence materials followed along inclined coronal loops towards the chromosphere,
and the funnel prominence disappeared. The AIA 171 \AA\  difference images reveal that the U-shaped coronal loops sunk down significantly as cool materials accumulated at the base of these loops
and rose up significantly as persistent cool materials draining from the base of the prominence. The descent of the U-shaped coronal loops and the formation of the V-shaped structure may 
suggest that cool materials transported into the coronal cavity dragged the magnetic field lines down, causing them to sag.  
Magnetic reconnection within or between  the sagging field lines may lead to the vertical drainage of prominence materials.
This observation demonstrates that chromospheric plasma transported into a coronal cavity and then drained out through vertical downflows
can result in the formation of a funnel prominence. 

The formed funnel prominence have a distinct horn-like structure (the U-shaped coronal loops) emanating from the prominence and extending into a coronal cavity
(see Figure 2({\it e} \sbond {\it i}) and Figure 3({\it a} \sbond {\it b})). 
\citet{gibson10} and \citet{karna15b} have indicated that the overall three-dimensional morphology of coronal prominence cavities
are long tubes with an elliptical cross-section. The long tubes may correspond to magnetic flux ropes made up of twisted magnetic field lines \citep{gibson06a,bak13}. 
Our observation suggests that the formed funnel prominence may be supported in a coronal cavity flux rope.
Therefore, the slow rise of the saddle-shaped coronal loops, and the gradually appearance of the U-shaped coronal loops and the coronal cavity could be well explained 
by slowly rotating a long flux rope whose axis distributed roughly along the solar latitude around the limb.
This observation fits very well with such a picture that chromospheric materials are transported into a coronal cavity by a rising cool column to form a funnel prominence.
The rising cool column, which might be a rising minifilament or an emerging magnetic flux rope carrying chromospheric materials,
could interact with the coronal cavity to transport its materials into the coronal cavity.
The scenario of rising minifilament is similar to the observations of  \citet{shen19} and \citet{yang19b},
while the scenario of magnetic flux rope emergence is consistent with the observations of \citet{oka10} and \citet{yang19a}.
However, the observations presented here are insufficient for a discussion of the carriers of chromospheric materials and the detailed interaction processes.
The chromospheric materials transported into the coronal cavity may drag the magnetic field lines downward and compress them at their base,
causing the magnetic field lines to sag  and further to form the V-shaped structure.
Afterwards, magnetic reconnection may occur within or between the sagging magnetic field lines,
resulting in the continued slide of cool materials from the V-shaped structure into the solar chromosphere,
followed by the rise of the U-shaped coronal loops and finally the formation of the funnel prominence.
To a certain extent, our observations reveal the interaction process between the transported prominence materials and the prominence magnetic fields.
In our observations, the persistent vertical drainage of cool materials from the base of the prominence to the solar chromosphere may be indicative of cross-field
slippage of cool, poorly ionized prominence material, perhaps involving magnetic reconnection \citep{low12a,low12b}.
Furthermore, these observations support the previous idea \citep{chae08,low12a} that the formed funnel prominence,
appearing as a vertical absorption thread on the AIA 193  \AA\ observation, might be supported in a vertical stack of magnetic dips of sagged coronal field lines.

Horn-like structures extend from the cool and dense prominence, connecting it to the hot and tenuous corona.
It is widely accepted that  horns are a part of the formation process of prominences from cooling and condensing coronal plasmas \citep{ber12,liu12,luna12,sch13,xia14}.
An opposite scenario was proposed by \citet{wang16} that horns are formed due to  the prominence evaporation,
likely involving some heating and diluting process of the central prominence mass. In the present study,
even though the appearance of the horn-like structures (the U-shaped coronal loops) could be attributed to the rotation of a coronal cavity flux rope into a favorable viewing angle,
 the interplay of prominence materials and magnetic fields may also play a role in the evolution of the horn-like structures.
 The chromospheric materials were transported into the coronal cavity and then they dragged the magnetic field lines downward, leading them to be compressed (see Figure 3({\it a} \sbond {\it b})).
As a result, the density at the base of the dragged and compressed U-shaped coronal loops may rise, intensifying the local emission 
and making the horn-like structures more visible (see Figure 2({\it e} \sbond {\it i}) and Figure 3({\it a} \sbond {\it b})).

Direct observational evidence for magnetic reconnection within or between the sagging magnetic field lines has not yet been found.
In the observations reported here, the descent of the U-shaped coronal loops and the formed V-shaped structure are direct observational evidence
that the U-shaped coronal loops are dragged by the cool materials accumulated in the coronal cavity to sag and generate deeper magnetic dips.
The projected velocities of the vertical downflows from the V-shaped structure to the chromosphere reach up to tens of kilometers per second,
which are commensurate with the velocities of the vertical downflows created in the numerical simulations by the magnetic reconnection between the sagging magnetic field lines \citep{kep14,jenk21}.
It is worthwhile to mention that the cross-field diffusion of neutral atoms in the solar prominence along the direction of gravity can also create vertical downflows.
However, the velocities of the vertical downflows created by the cross-field diffusion of neutral atoms are substantially lower than those found in the present study.
\citet{gil02} calculated the diffusion of neutral atoms in a simple prominence model and found that the magnitudes of the \ion{H}{0} and \ion{He}{0} downflow velocities
have values of about 370 cm s$^{-1}$  and 8.1 $\times$ 10$^{3}$ cm s$^{-1}$, respectively, for a density of 10$^{10}$ cm$^{-3}$.
In the course of the prominence materials sliding vertically into the chromosphere, the U-shaped coronal loops were evidently raised (see Figure 4({\it d})).
The vertical downflows, which might be caused by magnetic reconnection in or between sagging magnetic field lines, can lead to prominence material loss,
resulting in an upward magnetic tension force that exceeds gravitational force, forcing the U-shaped coronal loops to rise.
It is very likely that the persistent vertical downflows and the associated rise of the U-shaped coronal loops
are indirect evidence of magnetic reconnection in or between the sagging magnetic field lines.

This observation has far-reaching consequences for our understanding of prominence disappearance and eruptions.
Theoretical models of mass loading in eruptions \citep{fong02,low03} have predicted that prominence weight may play a role in storing magnetic energy for driving coronal mass ejections,
and significant mass-loss to a prominence may facilitate its eruption.
Some previous observations have shown that a sizable volume of filament material
drains along the filament axis toward their endpoints prior to its violent eruption \citep{bi14,jenk18,dai21}.
They suggested that the material drainage may be able to influence the local and global properties of the filament magnetic fields and thus facilitate its eruption.
However, what causes the filament material to drain along the filament axis towards its endpoints is still unknown.
The filament may be disturbed by its nearby eruptive activities and thus drive material drainage \citep{dai21}.
\citet{song23} found that the interaction between the filament and its nearby network magnetic fields could also cause material drainage.
In prominences, it is possible that the heavy prominence material in the magnetic dips may drag the prominence magnetic fields to sag
and drive the magnetic reconnection within or between these sagging magnetic field lines to form vertical downflows that appear as dynamic vertical threads \citep{chae10,low12b}.
On the one hand, the magnetic reconnection may modify the local magnetic field topology of the prominence.
On the other hand, material in the magnetic dips is considerably lost by the vertical downflows,
thereby reducing the gravity of the prominence material and accordingly increasing the net upward force, leading the prominence material to be lifted and drained along its magnetic field lines.
As a result, prominences may be  facilitated to erupt.
Further investigations are needed to determine whether gravity-driven vertical downflows inside prominences can cause prominence material to drain along the prominence magnetic fields,
leading to the disappearance or eruption of the prominence.

\begin{acknowledgments}
The authors sincerely thank the anonymous referee for detailed comments and useful suggestions for improving this manuscript.
The authors thank the scientific/engineering team of \emph{SDO} and GONG for providing the excellent data.
This work is supported by the National Key R\&D Program of China (2019YFA0405000);
the Natural Science Foundation of China, under grants 12073072,
12173084, 12273108, 12273106, 12203097, and 11933009;
the Yunnan Province XingDian Talent Support Program;  the ``Yunnan Revitalization Talent Support Program" Innovation Team Project (202405AS350012);
the CAS grant ``QYZDJ-SSW-SLH012"; and the Yunnan Key Laboratory of Solar Physics and Space Science (202205AG070009).
\end{acknowledgments}

\newpage
\begin{figure}
\epsscale{0.8}
\plotone{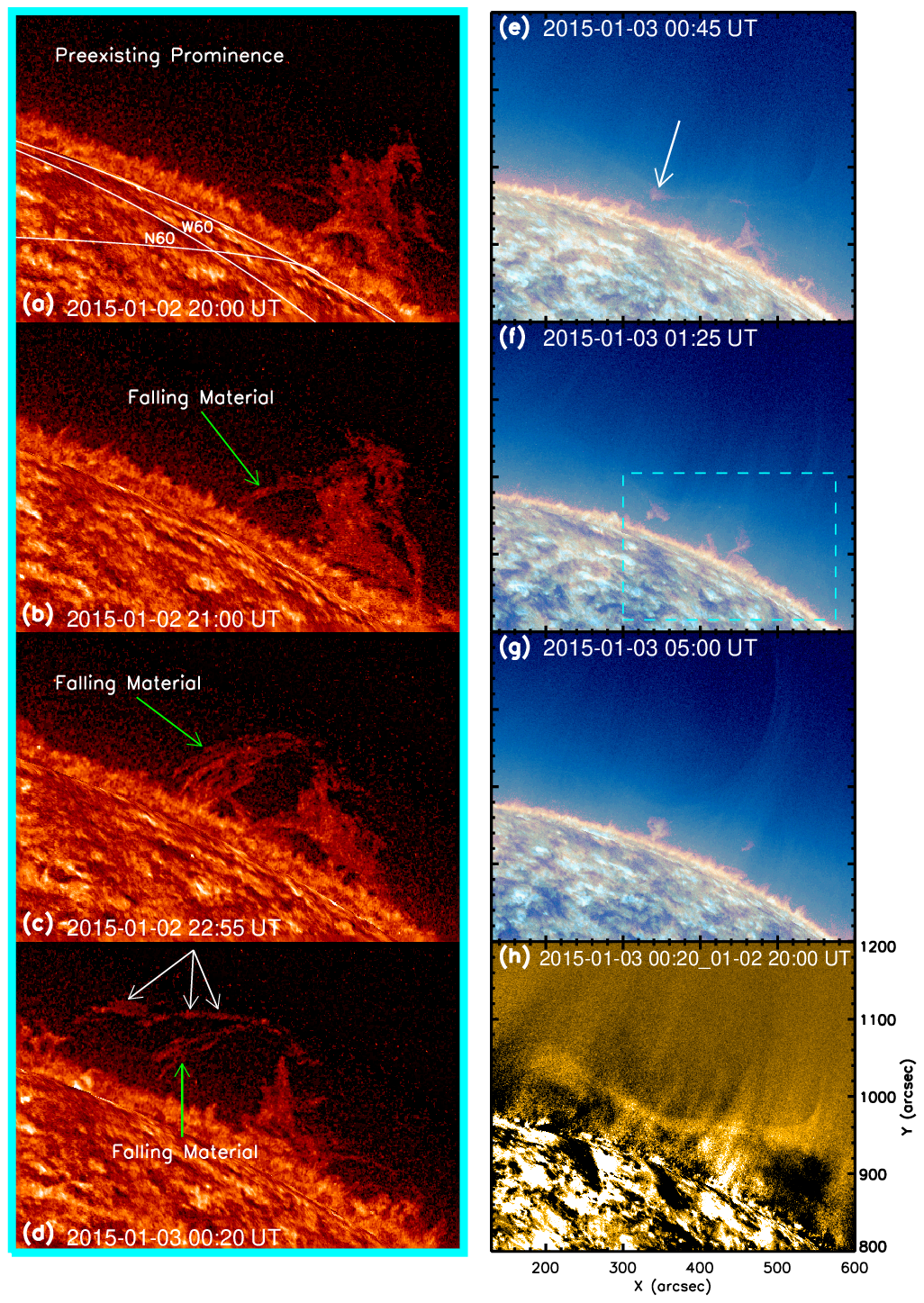}
\caption{Sequence of AIA 304 \AA\ (panels ({\it a} \sbond {\it d})) and composite of the AIA 304 \AA\ (red) and 171 \AA\ (blue) (panels ({\it e} \sbond {\it g})) images
showing the disappearance of a preexisting prominence and the formation of a new one. ({\it h}) An AIA 171 \AA\ difference image displaying the saddle-shaped coronal loops.
The white arrows in panels ({\it d}) and ({\it e}) point to the prominence material that moves along the saddle-shaped loops and sinks to their base.
The dashed box outlines the FOV of panels ({\it a} \sbond {\it d}).
An animation of panels ({\it e}) \sbond ({\it g}) is available. The animation has a 7 s cadence, covering 19:59 UT on 2015 January 2 to 05:02 UT on 2015 January 3.}
\end{figure}

\begin{figure}
\epsscale{1.}
\plotone{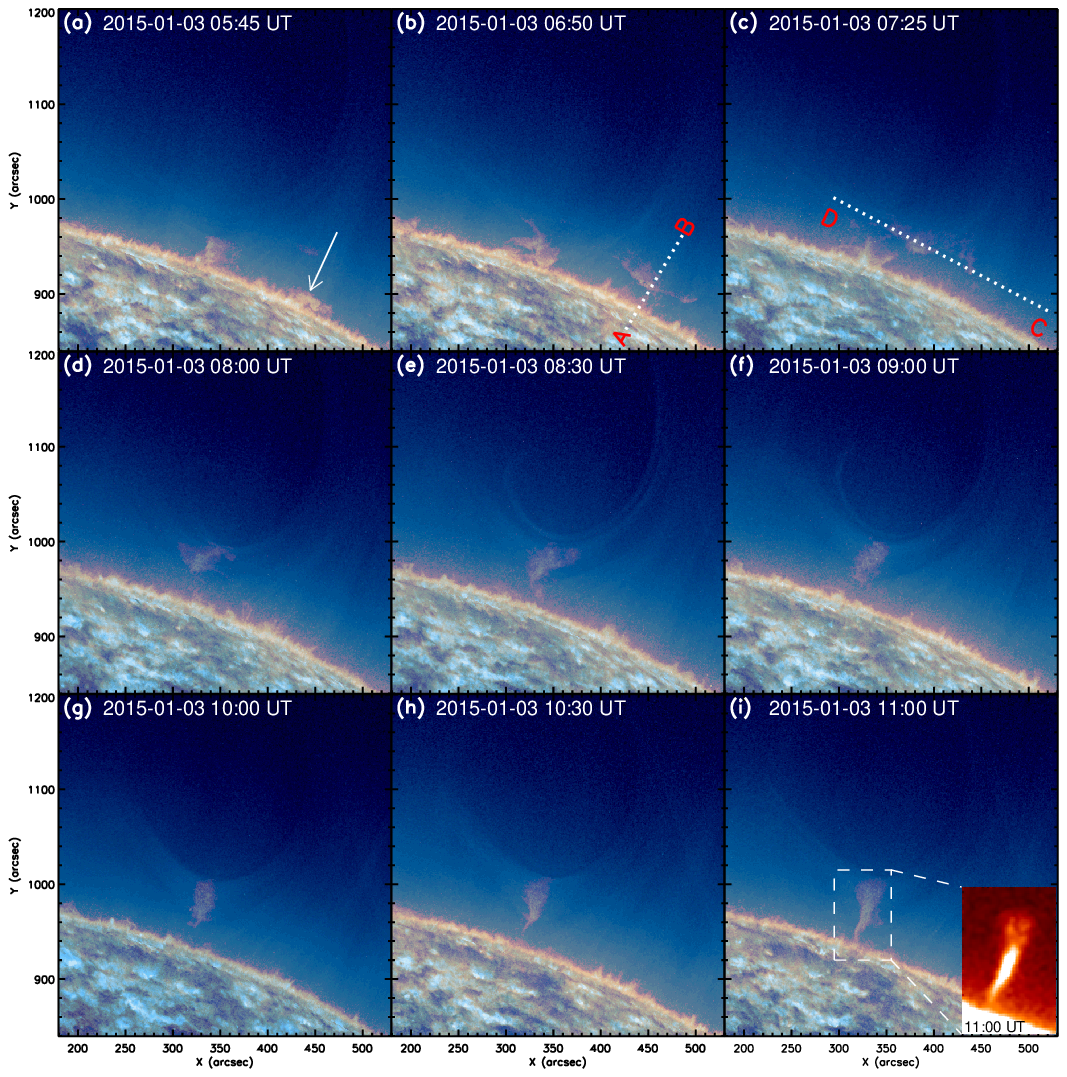}
\caption{({\it a} \sbond {\it i}) Composite of the \emph{SDO}/AIA 304 \AA\ (red) and 171 \AA\ (blue) images
display the formation of the funnel prominence and associated horn-cavity structures.
The white arrow in panel ({\it a}) points to a rising cool column.
The cool column is raised in the direction of the dotted line, ``AB'', and then is transported into the U-shaped coronal loops in the direction of the dotted line, ``CD''.
The dotted lines, AB and CD also mark two slits and are used for making the distance-time plots as shown in Figure 3. The inserted image is a GONG $H_{\alpha}$ image.
The formed funnel prominence,  with bright horn-like structures emanating from it and extending into a coronal cavity,
shows an inverted cone structure pointing sunward.
An animation of this figure is available. The animation has an 8 s cadence, covering 05:00 UT to 11:00 UT on 2015 January 3.}
\end{figure}

\begin{figure}
\epsscale{1.0}
\plotone{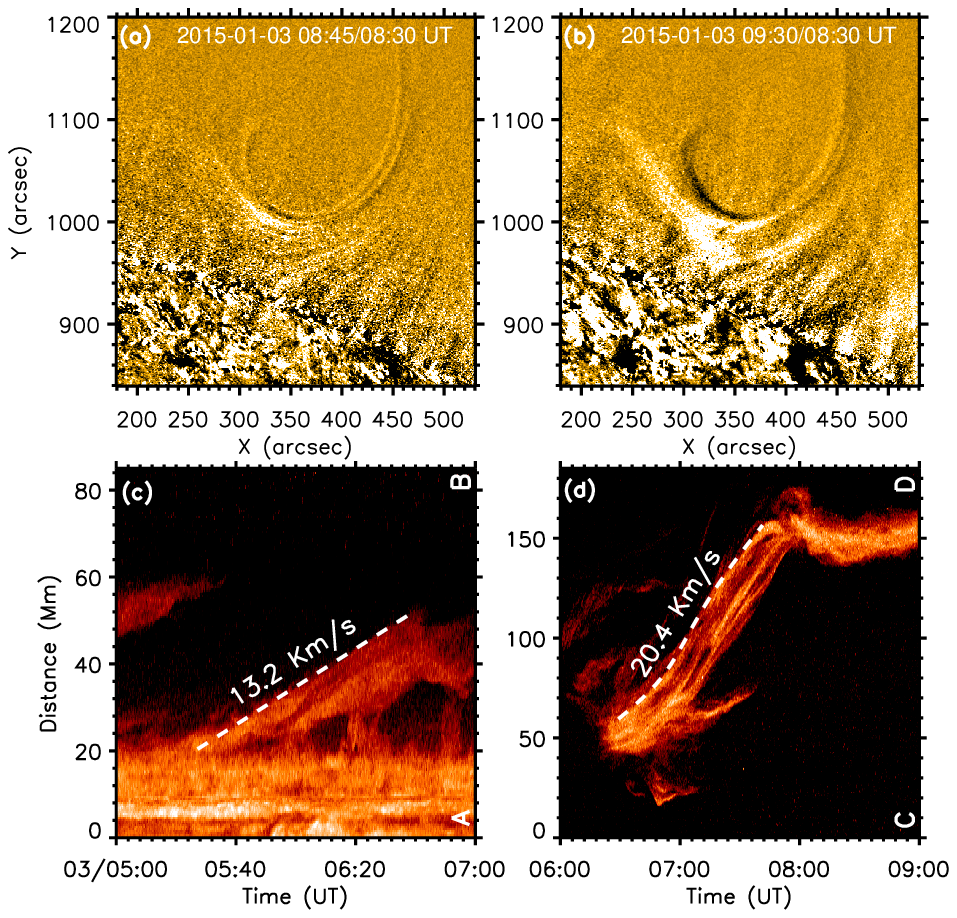}
\caption{({\it a} \sbond {\it b}) AIA 171 \AA\ difference images displaying the prominence materials drag the U-shaped coronal loops down,
leading to the compression of the U-shaped coronal loops at their base.
({\it c} \sbond {\it d}) Distance-time plots from AIA 304 \AA\ images for the two slits ( AB and CD) as shown in Figure 2({\it b}) and ({\it c}).
 The white dashed lines outline the motions of the rising cool column.}
\end{figure}

\begin{figure}
\epsscale{0.9}
\plotone{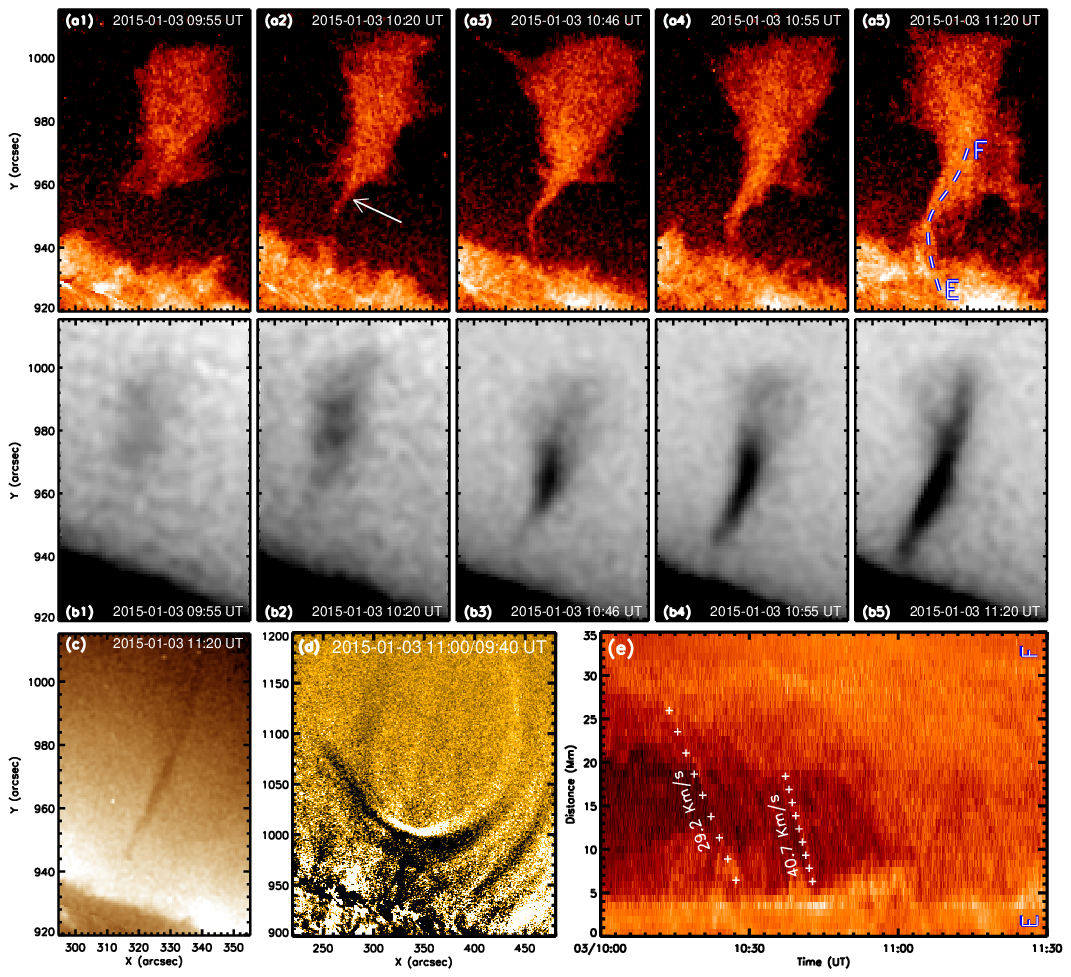}
\caption{Sequence of AIA 304 \AA\ (panels ({\it a1} \sbond {\it a5}))
and GONG $H_{\alpha}$ negative (panels ({\it b1} \sbond {\it b5})) images presenting the detailed process that cool materials drain from a V-shaped structure
down to the solar chromosphere to form the funnel prominence.
The funnel prominence corresponds to a vertical absorption thread in AIA 193 \AA\ (panel ({\it c})) observation.
An AIA 171 \AA\ difference image (panel ({\it d})) showing the rise of the U-shaped loops during the prominence material drainage.
The white arrow points to a V-shaped structure. The dashed line ``EF" indicates the slit position of the distance-time plot as shown in panel ({\it e}).
The plus signs outline the motions of the draining material.
An animation of panels ({\it a1}) \sbond ({\it a5}) is available. The animation has a 4 s cadence, covering 09:29 UT to 11:20 UT on 2015 January 3.}
\end{figure}

\begin{figure}
\epsscale{0.8}
\plotone{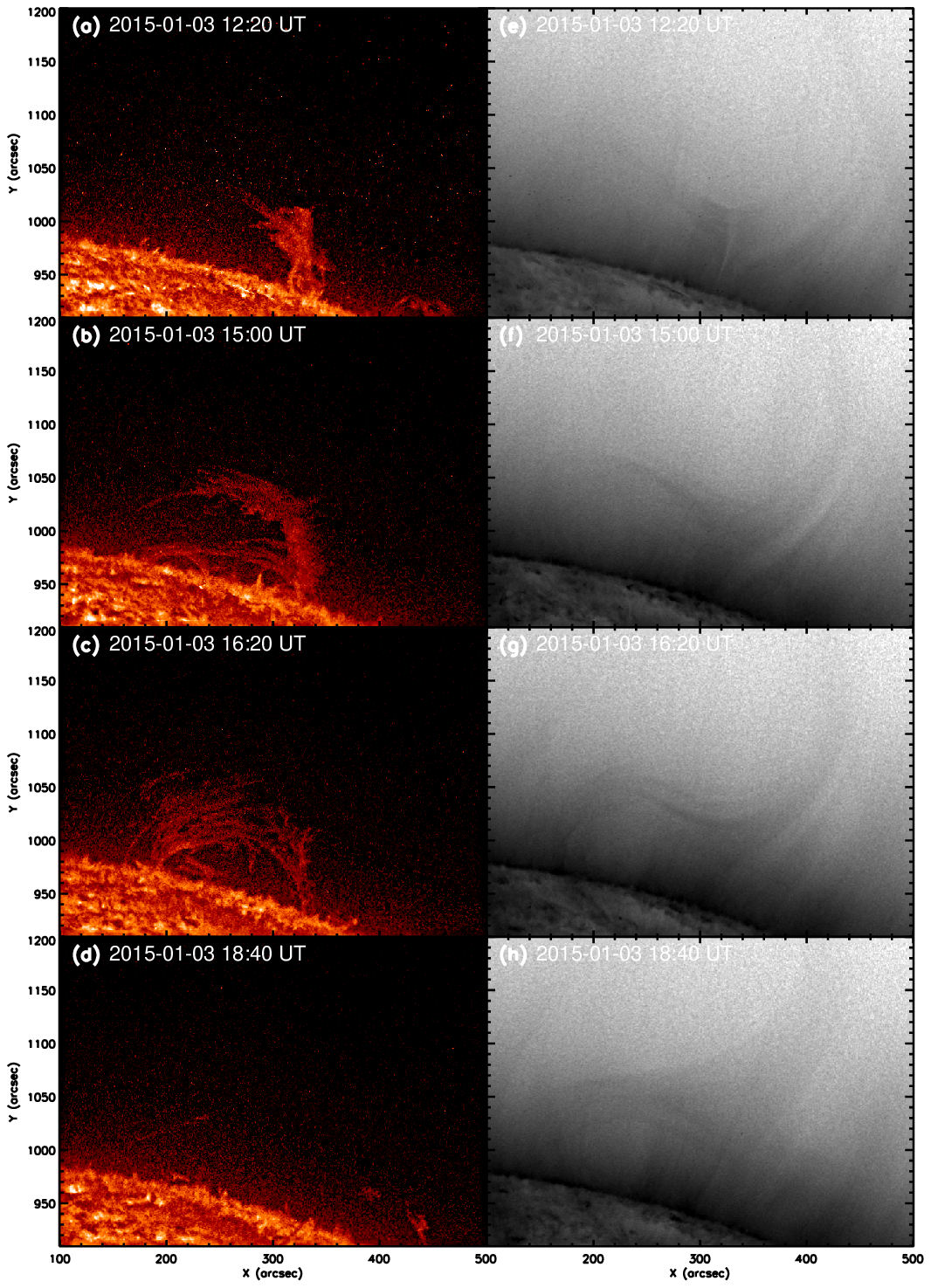}
\caption{AIA 304 \AA\ (panels ({\it a} \sbond {\it d})) and the negative 171 \AA\ (panels ({\it e} \sbond {\it h})) images
exhibiting the disappearance of the funnel prominence and the associated horn-like structure by the drainage of prominence material along well-defined curved coronal loops.
}
\end{figure}


\end{document}